ARTICLE

# Large area photoelectrodes based on hybrids of CNT fibres and ALD–grown TiO$_2$

A. Moya[a], N. Kemnade[b], M. R. Osorio[c], A. Cherevan[b], D. Granados[c], D. Eder[b]†, J.J. Vilatela[a]†

Hybridisation is a powerful strategy towards the next generation of multifunctional materials for environmental and sustainable energy applications. Here, we report a new inorganic/nanocarbon hybrid material prepared with atomically controlled deposition of a monocrystalline TiO$_2$ layer that conformally coats a macroscopic carbon nanotube (CNT) fiber. Through X-ray diffraction, Raman spectroscopy and photoemission spectroscopy we detect the formation of a covalent Ti-O-C bond at the TiO$_2$/CNT interface and a residual strain of approximately 0.7-2 %, which is tensile in TiO$_2$ and compressive in the CNT. It arises after deposition of the amorphous oxide onto the CNT surface previously functionalized by the oxygen plasma used in ALD. These features are observed in samples of different TiO$_2$ thickness, in the range from 10 to 80 nm. Ultraviolet photoemission spectroscopy on a 20 nm-thick TiO$_2$ coated sample gives a work function of 4.27 eV, between that of TiO$_2$ (4.23 eV) and the CNT fiber (4.41 eV), and the presence of new interband gap states that shift the valence band maximum to 1.05 eV below the Fermi level. Photoelectrochemical measurements demonstrate electron transfer from TiO$_2$ to the CNT fiber network under UV irradiation. Electrochemical impedance spectroscopy measurements reveal a low resistance for charge transfer and transport, as well as a large capacitance. Our results point to the fact that these hybrids, in which each phase has nanometric thickness and the "current collector" is integrated into the material, are very different from conventional electrodes and can provide a number of superior properties.

## Introduction

Hybrid materials based on metal oxides and nanocarbons have generated considerable attention for a wide range of applications such as photocatalysis,[1–4] dye-sensitised solar cells (DSSCs),[5–7] sensors,[8] batteries[9–11] and supercapacitors.[12–16] The large surface area of carbon nanotubes (CNTs), for example, enables the formation of large interfacial area with the inorganic coating and the possibility to form nanostructures with well controlled interfaces.[17]

Of the many available inorganic materials for energy-related applications, TiO$_2$ is very attractive due to the possibility of obtaining a large number of morphologies and controlling reaction sites, combined with a low cost and high chemical stability.[18] Various hybrids consisting of TiO$_2$ interfaced with CNTs and other nanocarbons have been produced[15,16] and used as photoanodes for diverse photocatalytic reactions, as well as for other applications in energy management.[19–22] The presence of CNTs can increase surface area[23] and light absorption in the visible range, often leading to the formation of additional band gap states in the electronic structure of TiO$_2$.[24]

But the main hypothesis is that the nanocarbon helps separate photogenerated carriers in TiO$_2$ by acting as an electron sink, increasing reaction rates by a reduction in electron-hole recombination rate.[25,26] This mechanism is generically evoked in the literature to explain improvements in a wide range of photocatalytic reactions, such as CO$_2$ photoreduction,[27,28] hydrogen production[29] and other photocatalytic reactions.[30] Interfacial charge transfer/accumulation processes are identified as crucial to improve the material performance, but experimental evidence of charge transfer at the TiO$_2$/CNT interface is surprisingly scarce.[31–33] This is mainly due to the synthetic methods used so far, predominantly based on wet-chemical synthesis of hybrid nanoparticles in dispersion. While this embodiment is natural for catalytic reactions, charge transfer dynamics in these systems can only be indirectly probed by complex techniques such as photoluminesce (PL) or transient adsorption spectroscopy (TAS). [34–36] Alternatively, the hybrid nanoparticles in dispersion can be deposited onto a flat substrate using a binder and sintered to produce a nanoporous electrode [37–39]. However, this leads to a high resistance between nanoparticles,[5,40,41] masking interfacial charge transfer processes in the hybrid and reducing photoelectrochemical properties of the resulting nanoporous electrode.[6,42]

Clearly, there is a need for the design and fabrication of TiO$_2$/nanocarbon hybrid architectures that minimise carrier diffusion in the inorganic phase and reduce charge transfer resistance to the nanocarbon by close proximity of the two phases. Such ensemble would enable direct measurement of interfacial processes, while simultaneously leading to high photocatalytic efficiencies by virtue of a lower overall resistance for photoelectron collection.

The pre-assembly of nanocarbons (e.g. CNTs into porous membranes subsequent coating with metal oxides is a particularly attractive strategy that leads to macroscopic hybrid (photo)electrodes with the inorganic/nanocarbon phases in close contact. Importantly, the nanocarbon electrode is immo-

[a.] IMDEA Materials, Madrid, Spain. Email: juanjose.vilatela@imdea.org
[b.] Institut für Materialchemie, Technische Universität Wien, Vienna, Austria. Email: dominik.eder@tuwien.ac.at
[c.] IMDEA Nanoscience, Madrid, Spain
†Corresponding authors.
Electronic Supplementary Information (ESI) available: additional characterisation data, sample scheme, HRTEM images, XRD patterns, Raman spectra and calculations are detailed. See DOI: 10.1039/x0xx00000x

bilised as a continuous porous network, with associated benefits for electrodes, catalyst reuse/recovery, etc. In such arrangement, the CNTs act as scaffold for the inorganic phase, while also functioning as current collector, for example for the transmission of carriers generated through photo(electro)catalytic processes. Used as support material, the benefits of such as network of interconnected nanocarbons are their mechanically stability and high conductivity.

The desired reduction in interfacial charge transfer resistance also calls for the *in-situ* growth of the $TiO_2$ on the nanocarbon surface, particularly by vapour-phase processes that can enable complete coating of the CNT surface by the inorganic phase with minimum detriment to its electrical properties or recourse to chemical functionalisation[43,44]. Amongst vapour-phase growth techniques, atomic layer deposition (ALD) is inherently suitable to coat large irregular porous CNT networks and control the deposition of the inorganic precursor on the atomic level. Indeed[45,46,47].

In this study, we prepared a new type of inorganic/nanocarbon hybrid material based on macroscopic fibres of CNTs covered with conformal layers of $TiO_2$ produced by ALD. Using such CNT fibres leads to hybrids where the current collector is effectively built into the material and makes it mechanically robust. Furthermore, the use of CNT fibre enables control over CNT type, as well as the production of samples with a large $CNT/MO_x$ interface (1500 $cm^2/cm^2$ of sample electrode), thus extending the range of characterisation techniques available. We present evidence of a covalent interfacial bond and the presence of residual strain, which is compressive in the CNTs and tensile in $TiO_2$. We also determine electronic structure of the hybrid material using photoelectron spectroscopy. Finally, we demonstrate that these hybrids are ideal photoelectrochemical electrodes, enabling for example direct probing of photogenerated charge transfer by simple photocurrent measurements. Finally, we present electrochemical impedance spectroscopy data under dark and illumination conditions, and a corresponding equivalent circuit that captures various interfacial charge storage/transfer processes in the hybrid, including an interfacial charge transfer resistance several order of magnitude lower than previous reports.

In addition to providing new insights into metal oxide CNT hybrids, this work highlights the potential of hybrids based on CNT fibres, a scaffold that combines the electrochemical properties of a porous carbon, high-performance mechanical properties, electrical conductivity superior to steel, and which is ideally suited for ALD.

## Results and discussion

### Morphology and crystal structure of TiO₂/CNT fibre hybrid

ALD technique was used for the growth of $TiO_2$ directly on the CNT fibre surface. As Figure 1a shows, the CNT fibre (CNTf) film consists of entangled bundles of CNTs (each of ~40 nm in diameter) that form a unique macroscopic fibre structure with open porosity. Figure 1b and Figure S1 show that the ALD precursor penetrated through the CNTf pores and nucleated on the surface-exposed CNTf bundles. A successful coating (controlled in the range of 10-100nm of $TiO_2$ thickness) of the CNT fibre was achieved as confirmed by the formation of a uniform and conformal layer of $TiO_2$ (Figure S2). Minor agglomerates of metal oxide were formed as a consequence of particle nucleation on the $TiO_2$ layer, which suggest that the nucleation and conformal growth of the titanium precursor is more favourable on the CNT fibre surface. Due to the presence of porosity in the CNT fibre, the created interfacial area with the oxide is very large, approximately the total area of the CNT fibre in the electrode that is 7500$cm^2$, as estimated from the surface area of the fibre (260$m^2$/g) and its linear density (0.6g/km). Overall, the interfacial area is much higher than that formed by $TiO_2$ deposited on planar 5$cm^2$ substrates. However, our $TiO_2$/CNTf hybrid electrodes are very flat from the macroscopic point of view (see inset of Figure 1b).

We optimised the ALD process parameters until the obtained coating of the CNT bundles was conformal and homogeneous. The histogram of Figure 1c shows the diameter distribution of the CNTf bundles and those coated with 20nm of $TiO_2$ layer evaluated from the SEM images. The diameter distribution of the hybrid agrees well with the theoretical thickness of metal oxide, 20nm of radius in this case, and the bundle diameter ranged between 20 and 50nm.

However, these conditions led to an amorphous coating of titanium oxide as was confirmed by XRD (Figure S3). Thus, a post-heat treatment at 400°C for 1h was carried out to crystallise the metal oxide into anatase phase. Inert atmosphere (Ar) was used to prevent the degradation of the CNT fibre . The annealing treatment in Ar atmosphere did not changed the conformability of the $TiO_2$ layer on the CNT surface and furthermore, the CNT structure was preserved. A uniform nucleation of $TiO_2$ layer on the CNT fibre surface was further confirmed by TEM (Figure 1d) that demonstrates a strong adhesion between, in this specific case, a 20nm of $TiO_2$ layer and a CNT bundle. In general, uniform coatings of the CNT fibres were observed with the amorphous $TiO_2$ layers and preserved after the heat treatment, leading to a preferential growth of the inorganic crystallites on the CNT fibre surface rather than random particle agglomeration. In addition, no porosity in the $TiO_2$ layer was observed in TEM images (see Figure S4 for further evidences), which implies that the CNT fibre acts not only as support but also as nucleation agent.[3] Figure 1d also illustrates the FFT of the image and a HRTEM image of the inorganic coating. Both show that the annealing at 400°C for 1h was enough to crystallise the $TiO_2$ into anatase phase. Specifically, the (101) anatase plane that corresponds to an interplanar distance of 0.352nm was mainly observed, suggesting a single-crystalline nature of the oxide coating, very unusual for $TiO_2$. The CNT fibre can easily dissipate the heat of the nucleation and crystal growth during the annealing and favours the oriented crystallisation.[48–50] The result is that the synthesized hybrid can be considered as a bicontinuous structure with an intimate contact between the TiO$_2$ and the CNT bundles and, importantly, forming large interface that correspond to (101) anatase/(002) CNT. Importantly, the coated CNT bundles are still connected along the fibre structure and act as electron transport pathways. Therefore, the CNT fibre acts as an active template for the nucleation and growth of TiO$_2$ and later as current collector. These results are in agreement with recent studies on arrays of vertically-aligned CNTs covered by large monocrystalline TiO$_2$ domains grown by ALD[51].

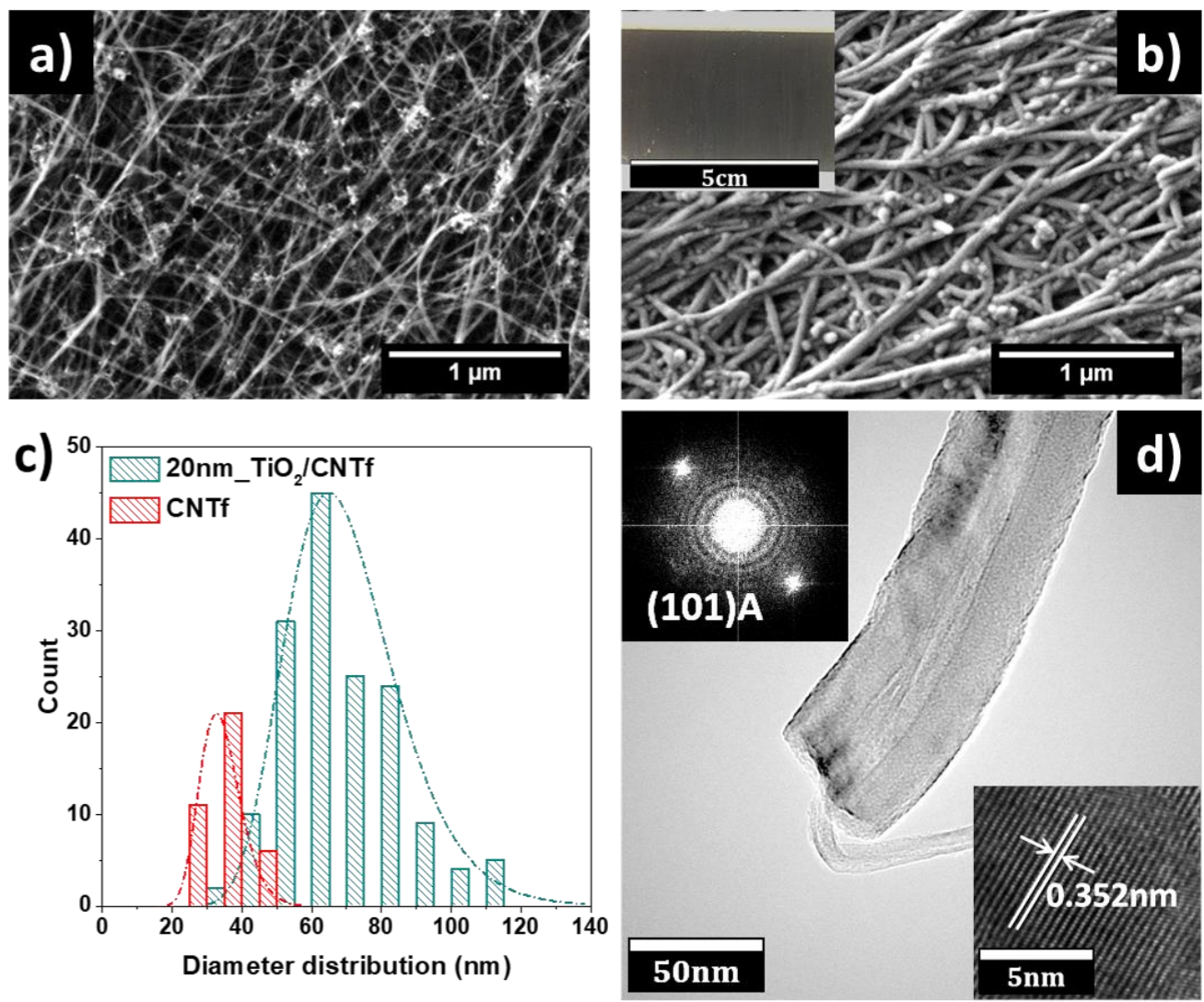

Figure 1: SEM image of a) CNT fiber film and b) TiO$_2$/CNT fiber hybrid after ALD deposition. c) Diameter distribution of CNTf bundles and a TiO$_2$/CNT fiber hybrid with the 20nm metal oxide layer. d) TEM image of TiO$_2$/CNT fiber hybrid annealed at 400ºC with a FFT and HRTEM showing the interplanar distance of the (101) anatase in the insets.

The crystal structure of the hybrid samples with different TiO$_2$ thickness and annealed at 400°C for 1 hour were analysed by XRD. The XRD patterns in the range of 2θ=23-29° of all the samples (Figure 2a) present pure anatase phase which is in agreement with TEM observations. Effectively, the annealing at 400°C for 1h crystallises the thin layer of metal oxide into pure anatase phase, whose diffraction peaks are more pronounced in the samples with the higher thickness (due to larger mass fraction of the oxide). The broadening of this peak is due to the overlapping contributions of the most intense peak of the anatase phase, the (101) plane at about 25.3°, and the interlayer spacing of the CNTs, the (002) plane at about 26.5° (see Table S1 for further detail).

Figure 2a also presents the deconvolution of this peak by fitting two Lorentzian curves. The change in position and width of the two peaks can be attributed to crystal size effects and/or microstrain induced by the inorganic growth on the CNT fibre surface. Possibly, the growth of the inorganic on the CNTf surface has pronounced effects on the crystal structure of both, TiO$_2$ and CNT fibre. The *in-situ* growth of the inorganic on the CNT surface produces a strong interaction between them, leading to a microstrain of 0.8-2%, especially relevant for the thinnest anatase coating since the interfacial area over the total analysed area is much higher for the samples with lower TiO$_2$ thickness. Also, such microstrain value is in the range determined for a graphene sheet when it is interfaced with (101) anatase surface.[52] The interaction between both components, for example through a chemical bond, causes disorder in the crystal lattice of the individual components, leading to a residual strain in the hybrid, which itself provides evidence of the strong interaction between the two phases.

In addition to XRD, Raman spectroscopy was used to analyse crystal structure as well as structural disorders in the $TiO_2$ originated during the *in-situ* growth on the CNT surface. In order to understand the different mechanisms acting during the formation of a CNTf/$TiO_2$ hybrid, Raman measurements were carried out on various samples, including those with different $TiO_2$ coating thickness, both before and after crystallisation, as well as CNTf exposed to the ALD oxidant but no $TiO_2$ precursors.

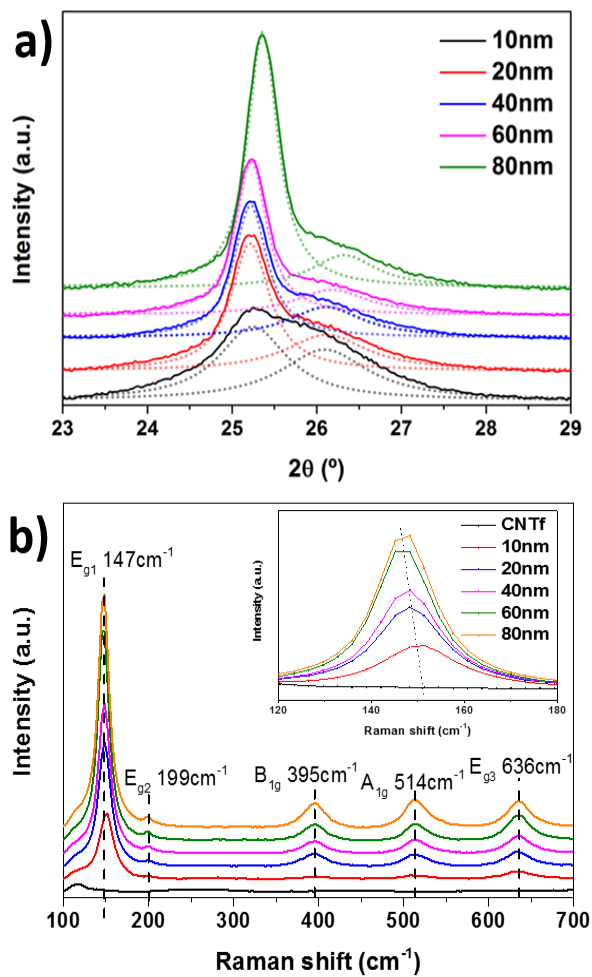

Figure 2: a) XRD patterns in the range of 23-29º showing the deconvolution of the anatase and CNT peaks for the hybrid samples with 10-80nm of $TiO_2$ thickness. b) Raman spectra of anatase active modes for the pure CNTf and $TiO_2$/CNTf hybrids. The inset shows a magnification of the most intense peak, where a shift to lower wavenumber is observed when the $TiO_2$ layer thickness increases.

Figure 2b shows the Raman spectra at the low wavenumber region of pristine CNT fibre and the crystalline $TiO_2$/CNTf hybrid samples containing different thickness of the oxide layer. The Raman spectra of the crystalline hybrid samples present the vibrational modes characteristic for $TiO_2$ anatase phase, which get more intense with increased $TiO_2$ thickness. The Raman bands at around 147 ($E_{g1}$), 199 ($E_{g2}$), 395 ($B_{1g}$), 514 ($A_{1g}$) and 636cm$^{-1}$ ($E_{g3}$), all correspond to anatase $TiO_2$ phase.[53] The most intense anatase peak, the $E_{g1}$ mode, shifts from 150cm$^{-1}$ down to 147 cm$^{-1}$ as $TiO_2$ thickness increases. Although the position of this mode is sensitive to factor such as crystal size and phonon confinement, these mechanisms are only significant for crystal sizes < 10nm.[54] Additionally, the direct evidence of crystal strain obtained by XRD strongly suggests that the observed Raman downshift is an indication of a tensile strain in $TiO_2$, which reaches its maximum at the interface with the CNTf and gradually decreases outwards from it. For reference, a -3 cm$^{-1}$ shift in this mode is equivalent to a hydrostatic pressure of 1GPa.[55] To a first approximation, this is equivalent to a strain around 0.7% (see calculations in SI), comparable to the strain measured by XRD.

Similarly, Raman spectroscopy was used to characterise changes in the CNT fibre upon hybridization and distinguish, for example, the effects of metal oxide deposition from those during the annealing stage. Such measurements take advantage of the high resonant Raman scattering of CNTs, particularly those of few layers.[56] Figure 3a presents the Raman spectra of pristine CNT fibre and annealed $TiO_2$/CNT fibre samples with 10 to 80 nm thick anatase layers, this time in the region of interest for CNTs. The main Raman active modes of CNTs are clearly observed: the G band presented around 1580cm$^{-1}$, the D band at around 1345cm$^{-1}$, the D´ at around 1620cm$^{-1}$, and the 2D band at around 2685cm$^{-1}$. The G band arises from tangential vibration of the hexagonal lattice of sp$^2$ carbon atoms. The D and D' modes indicate the presence of structural defects/disorder in the graphitic carbon structure, for example from sp$^3$ C bonding.[57] Commonly, the D/G intensity ratio is used to describe the quality of CNTs in terms of defects in the graphitic structure. The 2D band is a second order mode, highly sensitive to strain in the lattice, interaction between CNT layers and other effects.

The comparison between the CNT structure of as-grown fibre with that of the hybrids reveals a large increase of the D band intensity, with the D/G intensity ratio going from 0.17 to about 1.2 (Figure 3b). The increase in defects is also observed through the emergence of the D´ band in the hybrid samples. In order to investigate if the CNT fibre structure was modified due to the interaction with the titanium precursor or due to the oxidative atmosphere (300W oxygen plasma) during the ALD, a sample of CNT fibre was subjected to 5 minutes of $O_2$ plasma in the absence of precursors. Its Raman spectrum shows an abrupt increase of the D/G ratio to 0.91, confirming the strong oxidising effect of the plasma on the CNT fibre (Figure 3b).

When the sample is subjected to $O_2$ plasma cycles in the presence of the $TiO_2$ precursor (TDMAT) this leads to an even higher D/G ratio. However, we attribute this to the oxidized precursor acting on defects previously formed on the CNT fibre by the plasma. In fact, such defects act as functional groups that improve interaction with the metal oxide intermediate species and are ultimately a prerequisite for the growth of a conformal highly crystalline $TiO_2$ layers.

Although hybrid samples could be produced from purely thermal growth at 300°C, they showed clear evidence of poor interaction between the two phases (Figure S6). As an alternative to using a plasma to pre-functionalise the CNTs, O-containing precursors such as Titanium-isopropoxide (Ti(OCH(CH$_3$)$_2$)$_4$) can take this role as well as that of forming a metal oxide[51]

It is also worth noting that the annealing stage does not change the Raman spectrum substantially (Figure S5). This indicates that no further defects are introduced into the graphitic lattice and that the main coordination has already been established with the amorphous metal oxide.

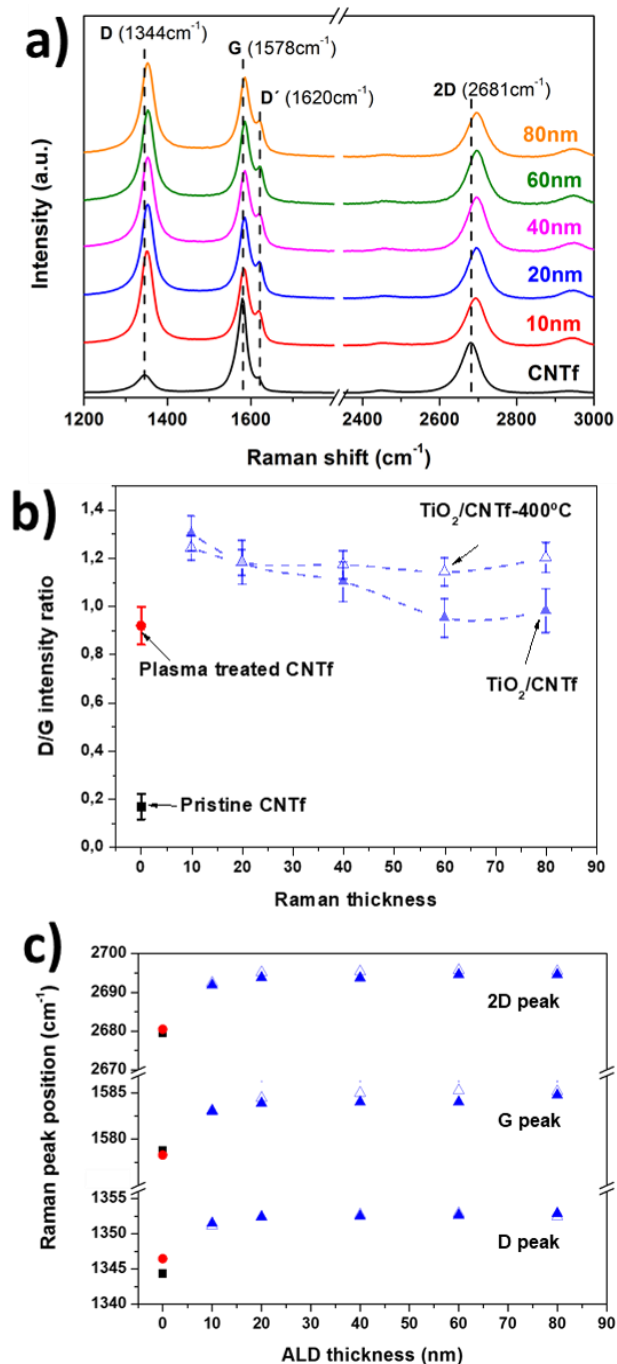

Figure 3: a) Raman spectra of TiO$_2$/CNTf hybrids after annealing at 400ºC for 1h in Ar atmosphere. b) D/G intensity ratio and c) shift of the CNT Raman modes, both as a function of the TiO$_2$ thickness in as-made (open triangles) and annealed (filled triangles) TiO$_2$/CNTf hybrids. Pristine CNTf (black squares) and CNTf exposed only to O$_2$ plasma (red circles) are included for reference.

In addition to the increase of D/G ratio for the hybrid samples, the hybridisation with metal oxide produces a displacement of the hybrid spectra to higher wavenumbers. The positions of the D, G, and 2D peaks for hybrids with different TiO$_2$ thickness are plotted in Figure 3c. It includes values for pristine fibres (black) and samples exposed to the O$_2$ plasma but no precursors (red). These two control samples show no changes in Raman mode frequency. The graph then shows a substantial upshift of the three Raman modes after the growth of the metal oxide on the CNTs. This occurs only for the first 20nm of TiO$_2$ coating, after which there is no substantial change in peak positions. The overall upshifts are large, around 7-9cm$^{-1}$, 4-7cm$^{-1}$ and 13-16cm$^{-1}$, for the D, G and 2D bands, respectively. A corresponding broadening of the peaks is also observed (Table S1). Interestingly, these changes occur after growth of the amorphous oxide, and no further shift is observed after annealing.

The data in Figure 3c show that the stiffening of CNT vibrational modes occurs as a consequence of the interaction with TiO$_2$. While doping effects cannot be entirely excluded,[58–60] we attribute the observed Raman downshifts mainly to a distortion of the graphitic lattice arising from the formation of a Ti-O-C bond[61–63] and the incoherent crystallographic registry between the two phases. Seen as deformation of the crystal, the Raman data can be conveniently used to calculate a strain (stress) in the CNTs at the interface with TiO$_2$. The 2D peak for single crystal graphite shifts at a rate of around -60cm$^{-1}$/ % tensile strain[64] and can in principle be used to calculate the corresponding axial strain. This requires knowledge of the Young's modulus of the material. In the samples used in this study, the TiO$_2$ covers *bundles* of nanotubes, rather than *individual* CNTs, and it is thus not trivial to determine precisely the strain of the CNTs at the interface with TiO$_2$. For the downshift measured here (16 cm$^{-1}$/%), the compressive strain is bound between 0.23%, assuming the modulus of an individual CNT (1TPa), and 0.4-3.2% based on *in-situ* Raman measurements during tensile testing.[65]

In summary, Raman data show that the CNTs are first functionalized by the oxidant (O$_2$ plasma) and then they couple strongly with the amorphous oxide through the formation of a covalent bond, resulting in a compressive axial strain. Subsequent crystallisation of the metal oxide produces small changes in the nanocarbon. Interestingly, this implies that the residual tensile strain in the first ≈10nm of the conformal TiO$_2$ layer is already present in the amorphous oxide.

**Interfacial chemistry and electronic structure of TiO$_2$/CNT fibre hybrid**

Further information of the chemical composition and chemical environment in the TiO$_2$/CNTf hybrid material was investigated by photoelectron spectroscopy for the hybrid sample with the 20nm-thick TiO$_2$. The XPS spectra were compared with those of the pristine CNT fibre and with pure equivalent 20nm layer of TiO$_2$ grown by ALD on a standard planar substrate (no CNT support). Figure 4: a shows the XPS spectra of the C1s region for the three samples. The intensity of the CNT fibre spectra is considerably higher than that of the hybrid or of the pure inorganic.

However, important differences can be observed between the spectra of the two latter components. Pure $TiO_2$ features two main peaks: the C-C signal at 284.8eV that corresponds to $sp^3$ hybridised carbon and a considerable contribution of C-O species at 287.6nm, associated with carbonate impurity species adsorbed on the $TiO_2$ surface. In contrast, the C1s spectra from the hybrid is similar to that of the pristine CNT fibre, which indicates that the changes in the C region are ascribed to the hybridisation process. Equipped with this information, the spectra can be analysed in more detail and the peak assignation can be made confidently.

Figure 4b presents details of the curve fitting applied to the spectra for pristine CNT fibres and the hybrid sample. The dominant contribution to the spectrum of the pristine CNT fibre is from the $sp^2$-hybridised C-C. There are small peaks from the $sp^3$-hybridised C-C as well as a small C-O signal. There is also evidence of π-π* interactions between CNT layers at 288-292eV[66], although overlapping with the O-C=O region. The highly graphitic nature of the pristine CNT fibres derived from XPS agrees with the low D/G intensity ratio observed in Raman.

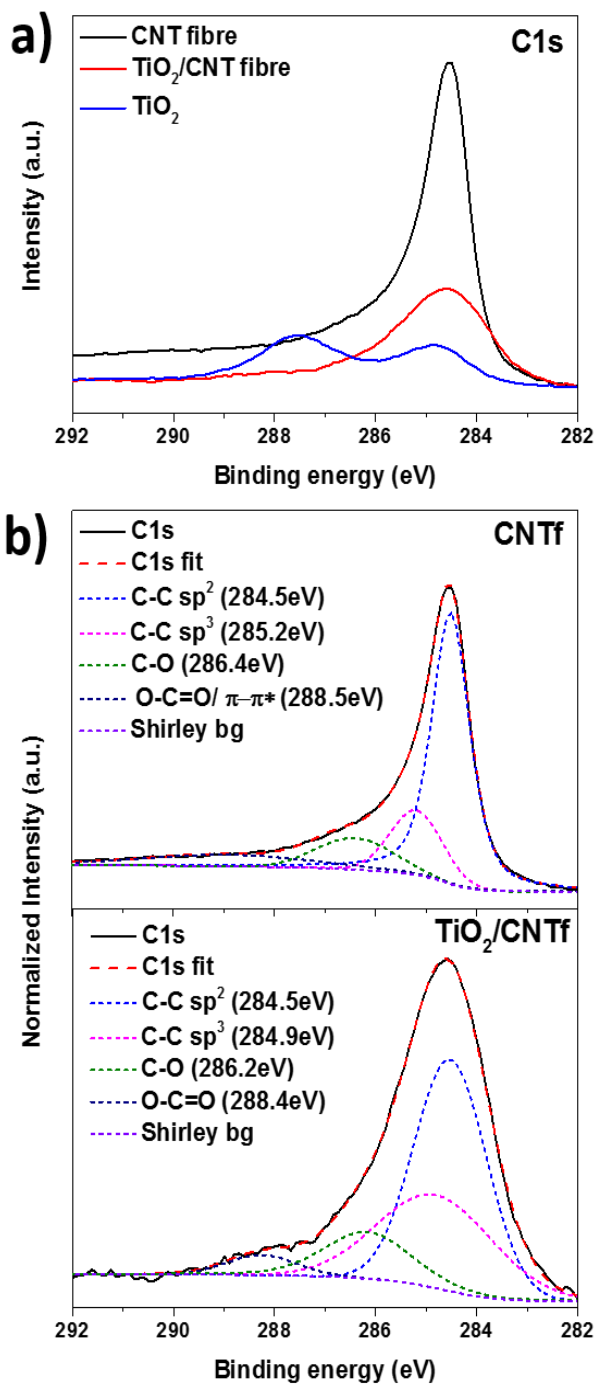

Figure 4: XPS spectra of pristine CNT fiber, $TiO_2$/CNT fiber hybrid and pure $TiO_2$ in the C1s region a) absolute spectra and b) fitted spectra for pristine CNTf (above) and $TiO_2$/CNTf hybrid (below).

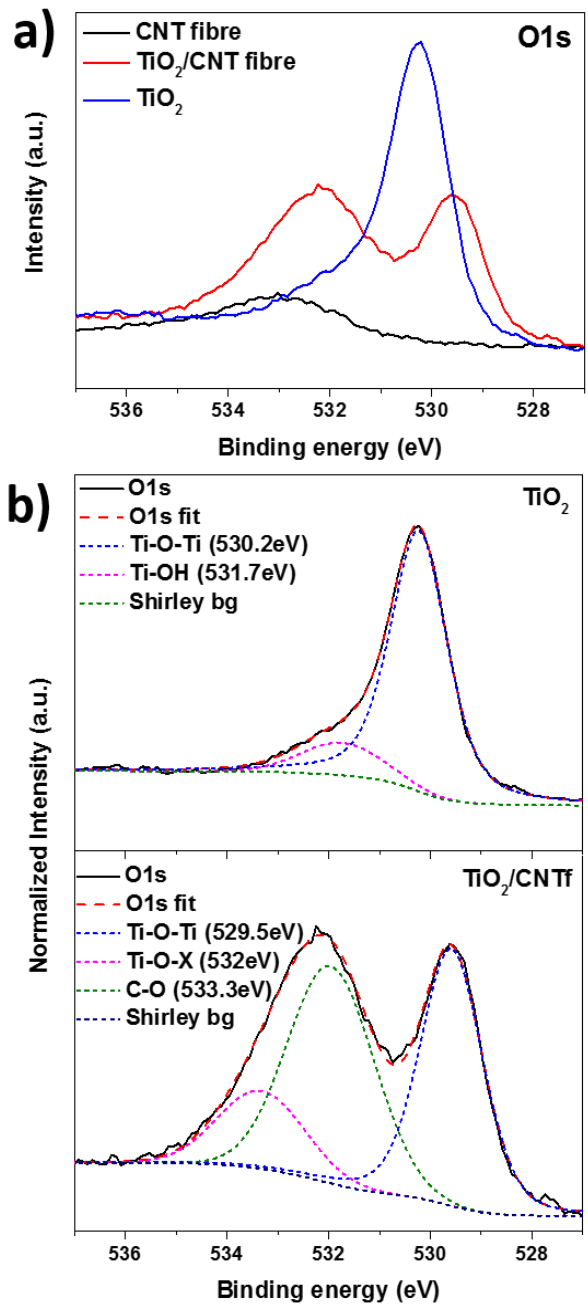

Figure 5: XPS spectra of pristine CNT fiber, $TiO_2$/CNT fiber hybrid and pure $TiO_2$ in the O1s region a) absolute spectra and b) fitted spectra for pristine $TiO_2$ (above) and $TiO_2$/CNTf hybrid (below).

In comparison, the C1s region for the hybrid shows an increases in the contribution of the $sp^3$-hybridised carbon, O-C=O and C-O species. There is a broadening of the $sp^2$-hybridised carbon peak, and reduction of the π-π* interaction. These observations are in good agreement with Raman observations.

Similarly, Figure 5 presents the O1s region of XPS spectra for the pristine CNT fibre and hybrid sample, where again, the direct comparison in absolute units facilitates signal assignment, for instance by confirming that the peaks in the hybrid spectra are not due to functional groups in the starting CNT fibre.

The pure $TiO_2$ reference presents a very clear peak at 530.2eV associated to Ti-O bonds and a shoulder at around 531.7eV ascribed to hydroxyl groups of the $TiO_2$ surface (Figure 5b). The hybrid spectrum contains three main features. The contribution from Ti-O bonds in $TiO_2$ gives a peak at 529.5eV, which is slightly shifted to lower binding energies in comparison with pure $TiO_2$, possibly due to charge transfer to O atoms favoured by the presence of C atoms bonded at the interface. The other two correspond to newly formed C-O (533.3eV) and Ti-O-X species (532eV).[67] The oxygen bonded with Ti and a heteroatom (Ti-O-X) is ascribed to an interfacial Ti-O-C bond between $TiO_2$ domains and the CNTs. This bond has been proposed in other $TiO_2$ nanocarbon systems,[4,67,68] reported recently for similar ALD-grown $TiO_2$ hybrids studied by XPS and STEM[51], and predicted by theory to have a formation energy of -1.74eV.[52] In combination with Raman data, XPS shows that the hybridization process involves the oxidation (functionalization) of the CNTs and subsequent formation of a covalent bond with the metal oxide. However, the large C-O peak (11,5%) in the hybrid suggests that not all functional groups bridge covalently the $TiO_2$/CNT interface.

In view of the strong interaction between the two phases in the hybrid, it is of great interest to study its electronic structure. Figure 6a illustrates the full UPS spectrum of the pure CNT fibre, $TiO_2$ and the same hybrid $TiO_2$/CNTf with 20 nm of ALD coating discussed above.

The $TiO_2$/CNTf hybrid exhibits an UPS spectrum similar to that of $TiO_2$, although more intense in the range of 5-12eV which corresponds to O and C states of the hybrid material. But more interestingly, to our surprise, the hybrid spectrum is appreciably more intense than pure $TiO_2$ at low biding energies. The inset in Figure 6a shows the valence band region, evidencing the much larger emission intensity for the hybrid sample. The spectrum is reminiscent of that of the CNTs, with a continuous density of states nearly up to the Fermi level. For the $TiO_2$ sample produced by ALD, the valence band maximum (VBM) is 2.6 eV with respect to the Fermi level, which is lower than the bulk band gap energy of anatase $TiO_2$ (~3.2eV) but in the range of reported values for oxygen-deficient $TiO_2$.[69] In contrast, the VBM for the hybrid is significantly smaller, at 1.05 eV below the Fermi level. All indications are that the samples have uniform coverage of the CNTs by $TiO_2$, which implies that the UPS signal is not due to a superposition of regions with and without $TiO_2$. Considering that the UPS signal comes from a small depth of the sample it is unclear how the electronic structure of the hybrid develops, and the distance over which the covalently attached CNTs has an effect on $TiO_2$.

Furthermore, the work function was determined by subtracting the onset of the secondary electron cut-off energy from the photon energy (21.2eV). No significant changes in the secondary edge were observed (see Figure 6b). The work function of the hybrid sample is calculated to be 4.27eV, an intermediate value between the CNT fibre (4.41eV) and the $TiO_2$ sample (4.23eV).

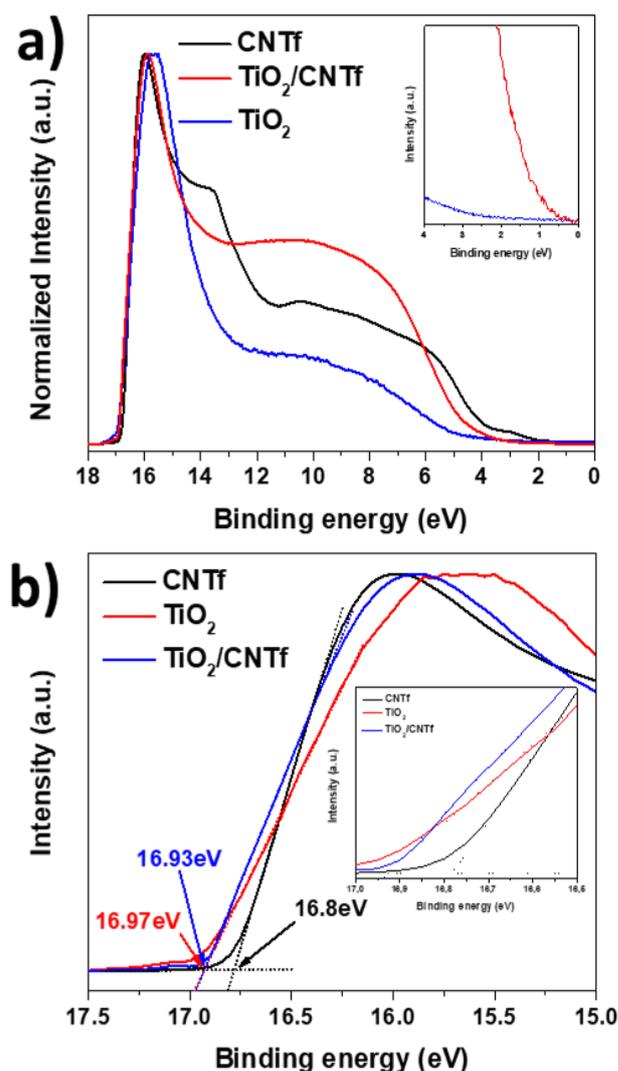

Figure 6: a) Full UPS spectra with inset of the low binding energy region of the hybrid compared with the $TiO_2$ and showing the increase of DOS in the valence band of the hybrid. b) Magnification of UPS spectra in the secondary electron cut-off for CNT fiber, pure $TiO_2$ and $TiO_2$/CNT fiber hybrid.

**Photoelectrochemical properties of $TiO_2$/CNT fibre hybrid**

The photoelectrochemical properties of $TiO_2$/CNTf hybrids with different $TiO_2$ thickness ranged between 10 and 80nm were tested in a three-electrode cell, with the hybrid used as a working electrode. A key feature of these samples is that they are already in the form of large-area nanoporous electrodes, which can be easily contacted (see Experimental details). The samples can also be peeld off the substrate to form flexible free-standing electrodes (Figure S7). But more importantly, their structure consists of two continuous phases, the highly conducting CNT network and the conformal $TiO_2$ crystalline layer. It provides a large semiconductor/liquid interface, while also minimizing diffusion length of photo(electro)generated carriers from $TiO_2$ to the CNT network current collector.

A first evidence of charge transfer in the hybrid materials is obtained by chronoamperometry measurements at zero bias potential. Figure 7a presents the photocurrent density for the different $TiO_2$/CNTf hybrids under UV light. Irradiation produces a rapid development of a positive photocurrent, indicating electron injection from $TiO_2$ to the CNT as in a photoanode. All the samples feature a quick response to the on/off cycles of the incident light and indicate strong dependence of the photocurrent density on the $TiO_2$ layer thickness. The photocurrent density reaches 2, 8, 16, 23 and 22µA/cm$^2$ for the $TiO_2$/CNTf samples with 10, 20, 40, 60 and 80nm, respectively. The photocurrent density increases with the $TiO_2$ layer thickness, reaching a maximum for 60nm, and then decreases. This is attributed to the absorption of $TiO_2$ in the UV range when the thickness increases, estimated in Figure S8.

The kinetics of charge transfer/storage at the hybrid electrode – liquid interface was studied by electrochemical impedance spectroscopy in dark and illumination conditions for the $TiO_2$/CNT fibre hybrid with 40 nm $TiO_2$ layer. Figure 7b shows the Nyquist plot in the dark and under irradiation. At open circuit potential conditions, the impedance spectra reveal an open semicircle with undefined time constants. The higher diameter arc corresponds to the dark conditions, whereas under UV irradiation, the impedance of the hybrid gets a well-defined capacitive arc and significantly reduced, suggesting more efficient charge separation.

The EIS data of the $TiO_2$/CNTf hybrid can be best fitted with the equivalent circuit (EC) model consisting of a resistance and two RC elements in series (Figure 7c). Other common circuits used to describe the properties of nanoporous electrodes did not fit the experimental data accurately.[70,71] The EC chosen has an obvious resemblance to the layered structure of the hybrid. In spite of its limitations it has been reported to accurately describe metal oxide thin films.[72] Briefly, the circuit consists of a series resistance (Rs) that contains all the external resistance contributions, such as the resistance of the electrolyte and the wires. The semiconductor/liquid interface is modelled by two double RC circuits in series. The first corresponds to the space charge layer ($Q_{sc}$) in the hybrid and the resistance of the $TiO_2$/CNTf hybrid interface ($R_{hybrid}$), which enables direct measurement of charge transferred at the interface between $TiO_2$ and CNTs. These processes correspond to the higher frequency component of the EIS data. The second RC element describes the double layer ($Q_{dl}$) and charge transfer resistance at the $TiO_2$/electrolyte interface ($R_{ct}$). Ionic diffusion and charge transfer processes at the semiconductor/liquid interface are slower and thus correspond to the lower frequency feature in EIS. The space charge layer capacitance has to be represented by a constant phase element, with its non-ideality probably arising from non-uniform distribution of charge in the hybrid.

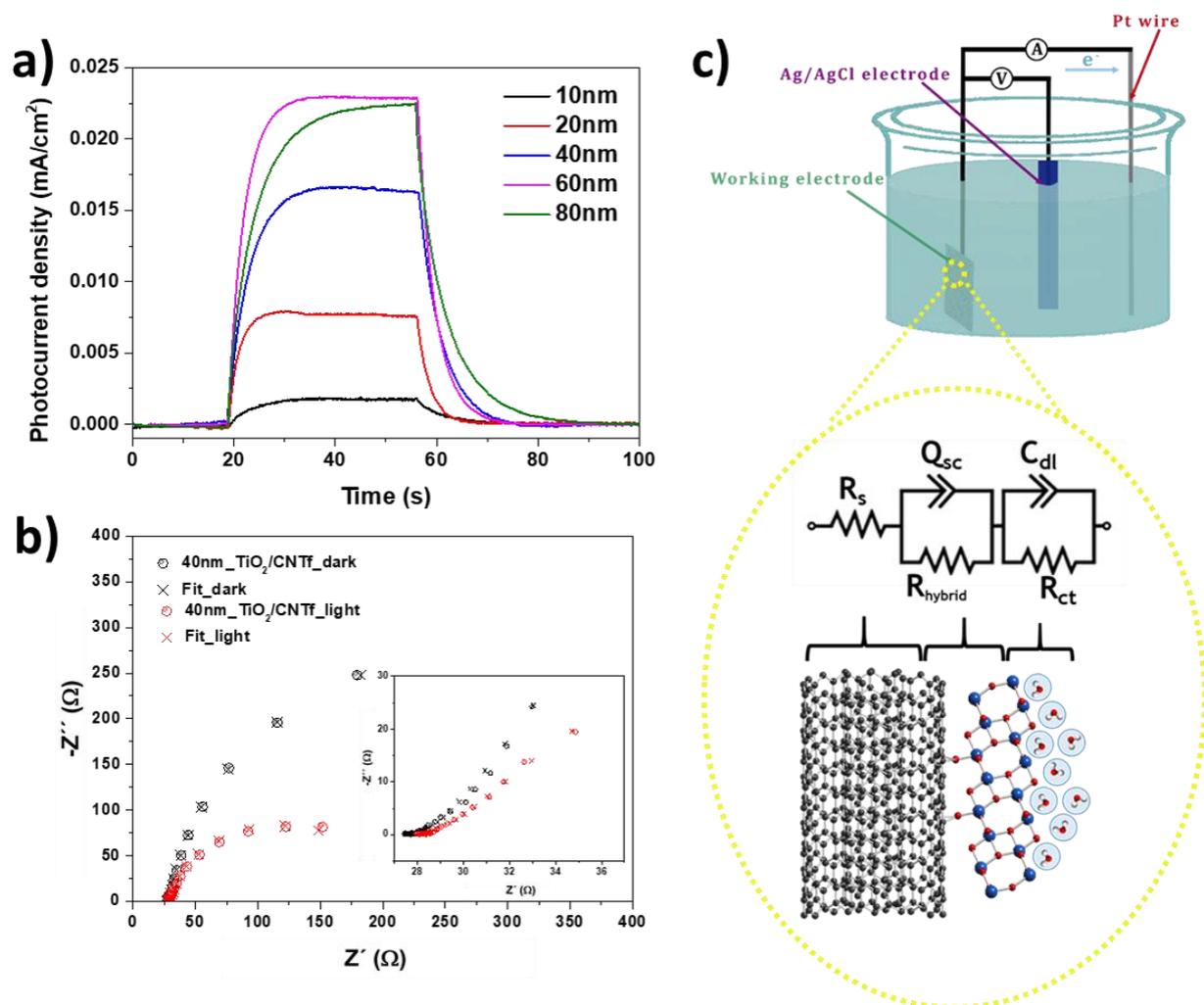

Figure 7: a) Chronoamperometry curves at open circuit potential of $TiO_2$/CNTf hybrid photoelectrodes. b) Nyquist plot of the $TiO_2$/CNTf hybrid and pristine CNT fiber under dark and illumination conditions. c) Representation of the three-electrode cell with a scheme of the $TiO_2$/CNTf hybrid as working electrode and the equivalent circuit model used for the fit of the impedance spectra.

The parameters obtained from the EC fit are shown in **Error! Reference source not found.**. The series resistance is low, at 28Ω under UV and in dark conditions. For the low frequency processes, the Helmholtz capacitance takes values of 51 and 68mF whereas the charge transfer resistance takes values of 514 and 135Ω for dark and light conditions, respectively. Effectively, the photogeneration of electrons in the conduction band of the semiconductor increases the photocarrier density at the semiconductor/electrolyte interface through charge separation and injection of electron from $TiO_2$ to the CNT, which is in agreement with the photocurrent measurements.

Table 1. Parameters obtained by fitting the EIS spectra of the $TiO_2$/CNTf hybrid photoelectrodes using the equivalent circuit $R_s+R_{hybrid}/Q_{acc}+R_{ct}/C_{dl}$.

|  | $Q_{sc}$ (F·s$^{a-1}$) | a | $R_{hybrid}$ (Ω) | $C_{dl}$ (F·s) | $R_{ct}$ (Ω) | $\chi^2/10^{-3}$ |
|---|---|---|---|---|---|---|
| **Dark** | 0.158 | 0.5 | 45.1 | 0.051 | 514 | 2.14 |
| **Light** | 0.144 | 0.5 | 10 | 0.068 | 135 | 1.63 |

At the high frequency range, the capacitance of the hybrid goes from 158 to 144mF in dark and illumination conditions, respectively. This semiconductor capacitance is higher than the Helmholtz capacitance, which is in contrast to the expected behaviour for a planar junction with the semiconductor thicker than the space charge layer. The large hybrid capacitance is tentatively attributed to the contribution from the chemical capacitance of the hybrid, with its components arising from surface and $TiO_2$/CNT interface states, as well as the chemical capacitance of the CNT current collector itself.[73] Finally, the hybrid resistance takes values of 45 and 10Ω for dark and light conditions, respectively. The very low resistance of the hybrid under UV irradiation implies a very low charge transfer resistance of photocarriers from $TiO_2$ to the CNT current collector. Even in dark conditions the relatively high current measured implies low transport and transfer resistance in the hybrid.

In Table 2 we compare values of resistance at the hybrid interface obtained in this work with the few literature values available for $TiO_2$/nanocarbon hybrid materials. We note that electrochemical measurement conditions and the choice of equivalent circuit differ; nevertheless, the hybrids introduced here show clearly a much lower interfacial charge transfer by several orders of magnitude. This highlights the benefits of the synthetic method used based on *in-situ* ALD growth of a thin conformal layer of metal oxide onto a pre-formed percolated CNT network.

Table 2. Comparison of charge transport/transfer resistance in $TiO_2$/nanocarbon systems.

| Materials | Resistance (Ohm) | | Ref. |
|---|---|---|---|
|  | Dark | Illumination |  |
| ALD $TiO_2$ on CNT fibres | 45 | 10 | This work |
| $TiO_2$/RGO (5%) | 223 x 10$^3$ | 6.2 x 10$^3$ | 40 |
| $TiO_2$/HCS | 581 x 10$^3$ | 4.2 x 10$^3$ | 74 |
| $TiO_2$/MWCNTs | n/a | 258.5 x 10$^3$ | 75 |

In summary, electrochemical measurements confirm the attractive properties of the new $TiO_2$/CNTf hybrids for their use as photoelectrodes. The control in terms of size, phase and composition of the inorganic growth on the nanocarbon surface has enabled the formation of a new $TiO_2$/CNTf hybrid material with high crystallinity and strong interaction between the two phases. The mild oxidation of the nanocarbon support promotes their the covalent interaction via a Ti-O-C bonds does not significantly disrupt the π-cloud of the nanocarbon. It keeps both components strongly coupled , facilitating photogenerated charge separation and transfer through the inorganic-carbon interface at atomic scale. The formation of inter-band gap states are likely to be crucial in the charge separation process and thus for the overall reduction in photocarrier recombination. In view of these results, we anticipate this new type of hybrids to have very high efficiency for photocatalytic oxidation/reduction reactions. Work is in progress to test this hypothesis.

## Conclusions

Thin layers of $TiO_2$ are grown *in-situ* on the surface of CNT fibres by atomic layer deposition. The $TiO_2$ coating is conformal and monocrystalline over large domains after annealing. The resulting structure consists of two continuous porous phases with a large common interface, approximately 7500cm$^2$, ideal to study interfacial processes. Through a combination of XRD, XPS and Raman spectroscopy, we observe a covalent Ti-O-C interfacial bond, and a residual strain of around 2%, which is compressive in the CNTs and tensile in $TiO_2$. Such strong interaction is first established after deposition of the amorphous metal oxide, before annealing crystallization, and is enabled by the modification of the $sp^2$ graphitic structure of the CNTf into a more chemically reactive surface ($sp^3$ hybridisation) formed by the $O_2$ plasma during the ALD process. These results highlight the need to chemically compatibilise the nanocarbon scaffold for subsequent integration of the metal oxide and which requires either gas-phase oxidation or the ALD metal oxide precursor to act as an oxidant. Considering the large effect such strain can have on the optoelectronic properties of the two phases in the hybrid, it is of interest to investigate this effect in more depth.

The hybrids have a work function of 4.27 eV, which is very close to the value of 4.23eV for a planar control sample of $TiO_2$ grown on a conventional flat substrate. To our surprise, UPS measurements indicate a high density of states in the hybrid at energies corresponding to the band gap of pure $TiO_2$, and which are reminiscent of the nanocarbon valence band. It is unclear how the small probing depth of UPS can access these features.

Photocurrent measurements demonstrate successful separation of photoexcited carries and injection of electrons from $TiO_2$ to the CNT conductive pathway. EIS measurements lead to an equivalent circuit consisting of two RC elements and a resistance in series. They confirm that both the $TiO_2$-CNT charge transfer and the transport across the hybrid have an associated low resistance. The large capacitance of the hybrid relative to the Helmoltz capacitance is attributed to the high chemical capacitance of the hybrid and its components. Work is in progress to confirm this hypothesis.

Overall, this work introduces a new type of large area porous hybrid material with high crystallinity of the two phases and the predominance of the large interface between them, leading to efficient charge separation. Our results point to the fact that these hybrids, in which each phase has nanometric thickness and the "current collector" is integrated into the material, are very different from conventional electrodes and can provide a number of superior properties in several applications.

## Experimental Section

**Materials**

CNT fibre was employed as macroscopic carbon material, which was produced by direct spinning method from the gas phase during the synthesis of CNTs by CVD.[66] For the atomic layer deposition, tetrakis(dimethylamido)titanium (IV) from Sigma-Aldrich was used without any modification as inorganic precursor.

**Preparation of hybrids:**

To produce $TiO_2$/CNTf hybrids, a film of CNTf was first deposited on a 5cm$^2$ glass substrate. Amorphous $TiO_2$ was grown by ALD (Fiji F200, Cambridge NanoTech, USA) deposition of the precursor (Ti(N(CH$_3$)$_2$)$_4$), injected in 0.1s pulses into the chamber, set at 135°C, followed by 35s of purging cycle with Ar and 20s of oxidising agent (oxygen plasma). The cycle was repeated for the formation of the desired $TiO_2$ thickness (10, 20, 40, 60 and 80nm), at a rate of 0.05nm/cycle. A smaller set of samples was grown at 150°C using water as oxidant in a Savannah 100 ALD Cambridge Nanotech system, with no structural differences observed. Because the ALD $TiO_2$ coating was amorphous after growth, the samples were subsequently annealed at 400°C for 1h in order to obtain crystalline anatase. The heat treatment was carried out in Ar atmosphere (100 L/min) to prevent damage of the CNT fibre. These conditions were selected based on our previous studies on electrospun CNT/$TiO_2$ hybrid samples [4].

**Characterization techniques:**

Morphological characterizations of the hybrids were carried out using a dual beam with field-emission scanning electron microscope (FIB-FEGSEM, Helios NanoLab 600i FEI) at 5 keV and high resolution transmission electron microscope (HRTEM, Talos F200X FEI) operating at 200 kV. Phase analysis was performed using X-Ray Diffraction (XRD), Empyrean, PANalytical using Cu Kα radiation. Raman spectra were obtained with a Renishaw PLC spectrometer with 532nm wavelength laser-excitation and 1 micron spot size. A laser power of 5.3mW was used, which we confirmed to be sufficiently low to avoid phase transformation. Peaks were fitted using mixed Gaussian/Lorentzian functions.

X-rays photoelectron spectroscopy (XPS) data were collected in a SPECS GmBH electron spectroscopy system provided with a PHOIBOS 150 +MCDanalyser. Ultraviolet photoelectron spectroscopy (UPS) were recorded using Thermo Scientific Multilab 2000 spectrometer with a 110 mm hemispherical sector analyser. UPS was measured with He I (21.2 eV) radiation source and total energy resolution of 200 meV. Energy scale was referenced to Fermi level measured on bare Au substrate.

**Electrochemical measurements**

The electrochemical measurements were carried out in a standard three-electrode cell with 0.1 M $Na_2SO_4$ electrolyte at pH 2 and using a Pt wire as counter electrode and Ag/AgCl reference electrode. The $TiO_2$/CNTf electrodes annealed at 400°C for 1h in Ar atmosphere were used as working electrodes with an active area of 5cm$^2$. The electrodes were electrically contacted by placing Cu foil on an 2cm$^2$ area that was masked during ALD and was therefore free of metal oxide The electrochemical measurements were recorded using a *SP-200 BioLogic* potentiostat under dark and illumination conditions (150W Xe lamp, λ<400nm). The power density of the UV light that reached the surface of the $TiO_2$ is 530mW/cm$^2$.

In order to study the electronic behaviour of the hybrid, the photocurrent density have been studied for the $TiO_2$/CNTf electrodes with different $TiO_2$ thickness using chronoamperometry technique at zero bias potential and intermittently illuminated with a manual chopper.

The kinetics of the charge transport and transfer processes were investigated by electrochemical impedance spectroscopy (EIS) for the hybrid photoelectrode with 40nm of $TiO_2$ layer thickness). Measurements were performed at zero bias potential under dark and UV illumination conditions, in the frequency range of 10kHz-0.1Hz and 10mV of signal modulation to avoid highly dispersive data and non-linear effects.

## Acknowledgements

JJV is grateful for generous financial support provided by the European Union Seventh Framework Program under grant agreements 678565 (ERC-STEM), by MINECO (RyC-2014-15115) and by Comunidad de Madrid (Spain) through the program S2013/MIT-3007 (MAD2D-CM) D.G. acknowledges RYC-2012-09864 and ESP2015–65597-C4-3-R.